\documentclass{ws-procs975x65-my} 

\usepackage{amssymb} 
\usepackage{amsfonts} 
\usepackage{amsmath} 
\usepackage{slashed}

  \usepackage{feynmp}

\newcommand{\beq}{\begin{equation}}
\newcommand{\eeq}{\end{equation}}
\newcommand{\beqa}{\begin{eqnarray}}
\newcommand{\eeqa}{\end{eqnarray}}

\newcommand{\half}{\frac{1}{2}}

\newcommand{\pbr}[2]{ \{ \hspace*{-2.6pt} [ #1 , #2\hspace*{1.4 pt} ] 
\hspace*{-2.6pt} \} }

\newcommand{\we}{\wedge}
\newcommand{\der}{\partial}

\newcommand{\ka}{\varkappa}

\newcommand{\Psib}{\overline{\Psi}}
\newcommand{\Phib}{\overline{\Phi}}

\newcommand{\what}[1]{\widehat{#1}}

\newcommand{\bx}{{\mathbf{x}}}

\DeclareMathOperator{\Tr}{Tr} 

\newcommand{\rd}{\mathrm{d}} 
\newcommand{\re}{\mathrm{e}} 
\newcommand{\ri}{\mathrm{i}} 

\begin{document} 

\unitlength = 1mm 

\title{On the ``spin connection foam" picture of quantum gravity \\ 
from precanonical quantization}
\author{Igor V. Kanatchikov$^*$}

\address{School of Physics and Astronomy, University of St Andrews,\\
St Andrews KY16 9SS, Scotland\\
$^*$E-mail: ik25@st-andrews.ac.uk \\ 
}

\begin{abstract} 
Precanonical quantization is based on a 
 generalization of the Hamiltonian formalism to field theory, the so-called De Donder--Weyl (DW) theory, which does~not~require~a~space\--time 
 splitting 
 and treats the space-time variables on an equal footing. Quantum dynamics is described by a precanonical wave function on the 
finite dimensional space of field coordinates and space-time coordinates, which satisfies a partial derivative precanonical Schr\"odinger equation. 
The standard QFT in the functional Schr\"odinger representation 
can be derived from the precanonical quantization in a limiting case.   
 An analysis of the constraints within the 
 DW Hamiltonian formulation of 
 the 
 Einstein-Palatini vielbein formulation of GR and quantization of 
  the 
  generalized Dirac brackets defined on differential forms 
 leads to the covariant precanonical Schr\"odinger equation for quantum gravity. 
 The resulting dynamics of quantum gravity is 
 described by 
 the wave function or transition amplitudes 
 on the 
 total space of the bundle of spin connections. 
 Thus, precanonical quantization leads to the 
 ``spin connection foam" picture of quantum geometry represented by a 
 generally non-Gaussian random field of spin connection coefficients, 
 whose probability distribution is given by the precanonical wave function. 
 The normalizability of precanonical wave functions 
is argued to lead 
 to the quantum-gravitational avoidance of curvature singularities. 
 Possible connections with LQG are briefly discussed.  
\end{abstract}

\bodymatter

\section{Precanonical quantization of fields} 
Contemporary quantum field theory originates from the canonical quantization  
which is based on the canonical Hamiltonian formalism. The latter dictates 
a picture of fields as infinite dimensional Hamiltonian systems. It also 
restricts the consideration to the globally hyperbolic space-times, as it 
implies a different role of the time dimension, along which the evolution proceeds,  and the space dimensions, which label the continuum of degrees of freedom of fields. Many problems we encounter in quantum gravity theories can be traced 
back to this very origin of QFT.

However, the canonical Hamiltonian formalism is not the only possibility to extend 
the Hamiltonian formalism from mechanics to field theory. The alternative ``Hamiltonizations" (i.e. writing the field equations in the first order form 
using some generalization of the Legendre transform) known in the calculus of variations \cite{kastrup} are, in fact, 
 inherently more geometrical 
 than the canonical formalism, 
 and they treat the space and time variables 
(i.e. the independent variables of the multiple integral variational problem) 
on the 
 equal footing, 
 i.e. essentially  as multidimensional 
 generalizations of the one-dimensional time parameter in mechanics. 

 In a sense, those Hamiltonizations of field theories are intermediate between the Lagrangean description and the canonical Hamiltonian formalism. Besides, in the case of mechanics, i.e. one-parameter variational problems, all those formulations reduce to the canonical Hamiltonian formalism in mechanics. 
For this reason we name those formulations ``precanonical" and the resulting 
procedure of quantization of fields ``precanonical quantization". 

The simplest precanonical formulation is the so-called De Donder--Weyl (DW) 
theory:\cite{kastrup}\    
for a Lagrangian density {$L = L(\phi^a, \phi^a_\mu, x^\nu)$,}  
which is a function of the field variables $\phi^a$, 
their first space-time derivatives {\small $\phi^a_\mu$,} 
and the space-time variables {$x^\mu$,} one defines the { polymomenta}
{$p^\mu_a := \frac{\der L}{\der \phi^a_\mu} $}
and the { DW Hamiltonian function} 
{$H(\phi^a, p^\mu_a, x^\mu) := \phi^a_\mu(\phi,p) p^\mu_a - L .$} 
Then, 
 if the Legendre transform $\phi^a_\nu \rightarrow p_a^\nu $ is regular, 
the 
field equations can be cast into the {DW Hamiltonian form}: 
\beq \label{dw}
\der_\mu \phi^a (x) = 
                            \frac{\der H}{\der p^\mu_a} , \quad 
\der_\mu p^\mu_a (x) = 
                                    - \frac{\der H}{\der \phi^a}   , 
\eeq 
This formulation requires 
neither a splitting into the space and time nor infinite-dimensional
spaces of field configurations.
The analogue of the extended configuration 
space here is a finite dimensional space~of~field variables~$\phi^a$ and 
space-time variables $x^\mu$, 
and the analogue of the extended phase space is a finite dimensional space of 
variables 
$p^\mu_a, \phi^a$ and $x^\mu$. Those spaces are bundles 
over  space-time (see e.g. Ref.~\refcite{geom-quant,deleon})  
whose 
sections are classical field configurations. 

While the usual approaches to 
field  quantization work on the infinite dimensional spaces of sections 
of the above bundles (``the spaces of solutions"), 
our precanonical quantization approach tries to find a formulation of 
quantum dynamics of fields in terms of the objects on the 
finite dimensional bundles themselves, without referring to 
the remnants of the classical thinking about fields as ``field configurations", 
similarly to the quantum mechanics, which has abandoned the notion 
of particles' trajectories in favor of the probabilistic concept of their 
location at different points of the configuration space 
or propagation amplitudes on this space. 

The Poisson brackets in the DW Hamiltonian 
 formulation of 
 field theory in $n$ space-time dimensions 
 are constructed\cite{my-pbr} using the  
  { polysymplectic $(n$+$1)$-form} 
on the extended polymomentum phase space 
 as the fundamental underlying structure generalizing the symplectic 2--form 
 of the canonical Hamiltonian formalism:  
$\Omega := 
dp^\mu_a\we d\phi^a\we \varpi_\mu$, 
where $\varpi_\mu:= 
 \imath_{{}_{\der_\mu}}\varpi, \; \varpi:=dx^0\!\we\!...\!\we\!dx^{n-1}$.  
The Poisson brackets are defined on differential forms 
representing 
  the local 
  dynamical variables (or ``observables") and  they lead to a Gerstenhaber algebra structure, which generalizes the standard Poisson algebra to the 
DW Hamiltonian formulation of field theory.\cite{my-pbr}

Precanonical quantization of fields\cite{my-quant} 
is based on quantization of the Poisson-Gerstenhaber 
brackets of forms\cite{my-pbr} in the DW Hamiltonian theory 
according to the Dirac quantization rule. 
It naturally leads to a description of quantum fields in terms of
{a~Clifford~(Dirac)~algebra~valued~wave~function~on~the~space~of~field~variables} and space-time variables, $\Psi (\phi^a, x^\mu)$, 
which fulfills 
 the 
 Dirac-like {precanonical Schr\"o\-dinger equation}\cite{my-quant} 
 with the DW Hamiltonian operator $\what{H}$ replacing~the~mass~term: 
\beq \label{nse}
\ri \hbar\varkappa \gamma^\mu\der_\mu\Psi = \what{H}\Psi,  
\eeq
where $\varkappa$ is an \mbox{ultraviolet constant of the dimension of the 
inverse spatial volume,}
{which originates from the representation of the dimensionful infinitesimal volume} 
 elements given by the differential forms $\varpi_\mu$ 
 in terms of the dimensionless Dirac matrices $\gamma^\mu$, 
and $\what{H}$ is a partial derivative operator with respect to the field variables. The natural appearance of Clifford algebra-valued functions and operators can be argued already on the 
level of geometric prequantization generalized to the DW Hamiltonian formalism.\cite{geom-quant} 
Let us note that the DW Hamiltonian equations (\ref{dw}) 
can be derived from (\ref{nse}) as the equations on 
 the expectation values of the 
corresponding precanonical quantum operators.\cite{ehrenfest} 
 The explicit form of a generalization of 
 equation (\ref{nse}) in quantum gravity will be presented below. 
 A validity of the Ehrenfest theorem in this formulation 
 of quantum gravity is 
 to be discussed elsewhere.\cite{ehrenhest-gr}
 
A relation of  the above description 
to the standard QFT in the functional Schr\"odinger representation  
 has been established in the limit of infinite $\varkappa$ or, more precisely, 
 when $\frac{1}{\varkappa} \gamma^0\!\rightarrow\!\varpi_0$.\cite{my-schrod}  
In this limiting case one can construct the Schr\"odinger wave functional 
as the multidimensional Volterra's product integral\cite{slavik} 
of precanonical wave functions and derive the canonical functional 
derivative Schr\"odinger equation 
from the precanonical  Schr\"odinger equation (\ref{nse}).\cite{my-schrod}  
Thus the standard QFT turns out to be a singular limiting case 
of the quantum theory 
of fields obtained via precanonical quantization. One can view the latter as 
an ``already regularized" quantum field theory in which the 
 UV scale $\varkappa$ is itself 
 an inherent element of the field quantization procedure, 
 that renders 
 standard cutoffs unnecessary.

\section{Precanonical quantization of vielbein gravity and the spin connection foam} 

A central role of the Dirac operator in precanonical quantization, 
which generalizes $i\der_t$ in quantum mechanics (cf. eq. (\ref{nse})), 
implies that gravity has to be quantized in vielbein formulation.
%
Here we follow our earlier work\cite{my-vielbein} 
(cf. Refs.~\refcite{my-metric} for an earlier work using the metric formulation).  

The Einstein-Palatini Lagrangian density with the cosmological term 
\beq \label{lagr}
{\mathfrak L}=  
\frac{1}{\kappa}
{\mathfrak e} e^{[\alpha}_I e^{\beta ]}_J 
\left(\der_\alpha \omega_\beta{}^{IJ} +\omega_\alpha {}^{IK}\omega_{\beta K}{}^J\right) + 
\frac{1}{\kappa}
\Lambda {\mathfrak e} ,
\eeq 
where 
the vielbein components $e^\mu_I$ and the 
 spin connection coefficients $\omega_\alpha^{IJ}$ 
 are  independent field variables 
 and ${\mathfrak e}:= (\det{\|e^\mu_I}\|)^{-1}$, 
leads to a singular DW Hamiltonian theory with 
 the primary constraints 
\beq \label{constr}
{\mathfrak p}{}^\alpha_{e_I^\beta}:=
\frac{\der {\mathfrak L} }{\der\, \der_\alpha e_I^\beta}
\approx 0 , 
\;\;  \mathrm{} \;\;  
{\mathfrak p}{}^\alpha_{\omega_\beta^{IJ}} :=
\frac{\der {\mathfrak L} }{\der\, \der_\alpha{\omega_\beta^{IJ}}}
\approx 
\frac{1}{\kappa}
{\mathfrak e} e^{[\alpha}_Ie^{\beta ]}_{J } .\
\eeq 
Those are second class, as it follows from 
the Poisson-Gerstenhaber brackets of  $(n$--$1)$-forms of constraints 
 $
\mathfrak{C}_{e^\beta_I}
:={\mathfrak p}_{e^\beta_I}^\alpha\varpi_\alpha 
\, \; \mathrm{and} \, 
\; 
\mathfrak{C}_{\omega_\beta^{IJ}}
:=  
\big( {\mathfrak p}{}^\alpha_{\omega_\beta^{IJ}} 
-
\mbox{$\ \mbox{$\frac{1}{\kappa}$}$}
{\mathfrak e} e^{[\alpha}_Ie^{\beta ]}_{J }\big) \varpi_\alpha 
$:
\beq \label{cbr}
 \pbr{\mathfrak{C}_e}{\mathfrak{C}_{e'}} =0 
 = \pbr{\mathfrak{C}_\omega}{\mathfrak{C}_{\omega'}} ,\; 
 \quad
 \pbr{\mathfrak{C}_{e^\gamma_K}}{\mathfrak{C}_{\omega_\beta^{IJ}}} 
 = - \mbox{\normalsize $\frac{1}{\kappa}$} \der_{e^\gamma_K} 
 \left( 
 {\mathfrak e} e^{[\alpha}_Ie^{\beta ]}_{J } 
 \right)  \varpi_\alpha  . 
\eeq
The DW Hamiltonian density obtained from (\ref{lagr}) reads 
\beq \label{hdw1}
\mathfrak{H}:= 
\frac{\der {\mathfrak{L}}}{\der\, \der_\alpha \omega}
\mathfrak{p}^\alpha_\omega   
+ 
\frac{\der \mathfrak{L}}{\der\, \der_\alpha e}
\mathfrak{p}{}^\alpha_e 
-  \mathfrak{L} 
= - 
{\mathfrak e} e^{[\alpha}_I e^{\beta ]}_J 
\omega_\alpha^{IK}\omega_{\beta K}^J 
- 
\frac{1}{\kappa}
\Lambda{\mathfrak e} .
\eeq

 Using a generalization 
of the 
 constraints analysis to the DW theory\cite{my-dirac}
we obtain an amazingly simple algebra of 
 the 
 fundamental Dirac brackets 
 on the subalgebra of $(n$--$1)$- and $0$-forms: 
\cite{my-vielbein} 
\vspace*{-5pt} 
\beqa \label{dbr11}
{}&\pbr{{\mathfrak p}^\alpha_\omega \varpi_\alpha}{\omega'\varpi_{\alpha'}}{\!}^{\rm D} 
 =\pbr{{\mathfrak p}^\alpha_\omega \varpi_\alpha}{\omega'\varpi_{\alpha'}}
 \,\,=\,\, \delta_\omega^{\omega'} \varpi_{\alpha'},
\label{dbr12} 
\\
{}&\!\!\hspace*{-10pt}
\pbr{{\mathfrak p}^\alpha_e \varpi_\alpha}{e' \varpi_{\alpha'}}{\!}^{\rm D}= 
 0  = %
\pbr{{\mathfrak p}^\alpha_e \varpi_\alpha}{ {\mathfrak p}_\omega}{\!}^{\rm D} 
\!=\pbr{{\mathfrak p}^\alpha_e \varpi_\alpha}{\omega' \varpi_{\alpha'}}{\!}^{\rm D} 
\!=\pbr{{\mathfrak p}^\alpha_\omega\varpi_\alpha}{e' \varpi_{\alpha'}}{\!}^{\rm D} 
, 
\label{dbr13}
\\ 
 \label{dbr11b}
{}&\pbr{{\mathfrak p}^\alpha_\omega \varpi_\alpha}{\omega'}{\!}^{\rm D} 
 =\pbr{{\mathfrak p}^\alpha_\omega \varpi_\alpha}{\omega'} \,\,=\,\, \delta_\omega^{\omega'},
\label{dbr12b} 
\\
{}&\!\!\hspace*{-10pt}
\pbr{{\mathfrak p}^\alpha_e \varpi_\alpha}{e'}{\!}^{\rm D}=0 = 
\pbr{{\mathfrak p}^\alpha_e \varpi_\alpha}{ {\mathfrak p}_\omega}{\!}^{\rm D} 
\!=\pbr{{\mathfrak p}^\alpha_e \varpi_\alpha}{\omega}{\!}^{\rm D} 
\!=\pbr{{\mathfrak p}^\alpha_\omega\varpi_\alpha}{e}{\!}^{\rm D}, 
 \label{dbr13b}
\\ 
\label{dbr11c}
{}&\pbr{p^\alpha_\omega}{\omega'\varpi_\beta}{\!}^{\rm D}=
\pbr{p^\alpha_\omega}{\omega'\varpi_\beta} = \delta^\alpha_\beta \delta^\omega_{\omega'},
\label{dbr12c} 
\\
{}&\!\!\hspace*{-10pt}
\pbr{{\mathfrak p}^\alpha_e}{e' \varpi_{\alpha'}}{\!}^{\rm D}=0=
\pbr{{\mathfrak p}^\alpha_e }{ {\mathfrak p}_\omega \varpi_{\alpha'}}{\!}^{\rm D} 
\!=\pbr{{\mathfrak p}^\alpha_e }{\omega \varpi_{\alpha'}}{\!}^{\rm D} 
\!=\pbr{{\mathfrak p}^\alpha_\omega}{e' \varpi_{\alpha'}}{\!}^{\rm D} \!=0 . \,\label{dbr13c}
\eeqa

The fundamental brackets are quantized 
 according to 
 the generalized Dirac's quantization rule:  
\beq \label{dirule}
[\hat{A}, \hat{B}]= 
- \ri\hbar \what{\mathfrak{e}\pbr{A}{\!B\!}{\!}}{}^{\rm D}, 
\eeq
in which  the presence of the operator~$\hat{\mathfrak{e}}$ 
 guarantees that tensor densities are  quantized as density-valued operators. 
  Quantization of fundamental Dirac brackets (\ref{dbr11})--(\ref{dbr13c}) 
  and  the equations of  constraints (\ref{constr}) 
  leads to the  representation of the operators of vielbeins: 
 $$\hat{e}{}^\beta_I = -\ri \hbar\ka\kappa \bar{\gamma}^{J}\frac{\der}{\der \omega_{\beta}^{IJ}},$$ 
and 
polymomenta of spin connection: 
$$\hat{{\mathfrak p}}{}^{\alpha}_{\omega_\beta^{IJ}} 
 = - \hbar^2\ka^2\kappa 
\,\hat{\mathfrak e}\,
\bar{\gamma}^{KL}\frac{\der}{\der \omega_{[\alpha}^{KL}}
\frac{\der}{\der \omega_{\beta]}^{IJ}},$$ 
 where 
$\bar{\gamma}^{J}$ 
are the fiducial 
 flat-space Dirac matrices. 
 Those Clifford-valued operators act on Clifford-valued  
  quantum gravitational precanonical wave functions  
 $\Psi=\Psi(\omega^{IJ}_\alpha, x^\mu)$  
 on the total space of the 
 configuration bundle of spin connections over the 
 space-time.\footnote{We use the notion of the ``bundle of connections" in 
 the sense similar to Ref.~\refcite{marco} 
 rather than e.g. Ref.~\refcite{viallet}. 
 }

We can also  construct the operator of DW Hamiltonian density 
$\mathfrak{H}=:{\mathfrak e}H$ restricted to the surface of 
constraints  $C$ given by  (\ref{constr}). 
 From (\ref{hdw1}) and (\ref{constr}) 
we obtain 
${({\mathfrak e} H)|_C} = 
- {\mathfrak p}{}^{\alpha}_{\omega_\beta^{IJ}} \omega_\alpha^{IK} \omega_{\beta K}{}^{J} 
- \frac{1}{\kappa} \Lambda{\mathfrak e}$ 
\ and, using the above representations,  
\begin{equation} \label{hgrop}
\what{H} = \hbar{}^2\ka^2\kappa\, \bar{\gamma}^{IJ} 
\omega_{[\alpha}{}^{KM}\omega_{\beta] M}{}^L 
\frac{\der}{\der \omega_{\alpha}^{IJ}} \frac{\der}{\der \omega_{\beta}^{KL}} 
- \frac{1}{\kappa} \Lambda .  
\end{equation}
Now, 
the covariant precanonical  Schr\"odinger equation for quantum gravity  
(cf. (\ref{nse}))
\beq \label{nsepsi}
\ri \hbar\ka 
\what{\slashed\nabla}  \Psi = 
\what{H} \hspace*{-0.0em} \Psi ,   
\eeq
where 
 $\what{\slashed\nabla} 
:=  
  \hat{\gamma\, }{\!}^\mu (\der_\mu+   
 \frac{1}{4} \omega_{\mu IJ} \bar{\gamma}^{IJ})$   
 and $
 \hat{\gamma\, }{\!}^\mu:= \bar{\gamma}{}^I \hat{e}{}^\mu_I   
 = - \ri \hbar\ka\kappa \bar{\gamma}^{IJ}\frac{\der}{\der \omega_{\mu}^{IJ}}$, 
can be written in an explicit form 
(up to an ordering of $\omega$-s and $\der_\omega$-s):   
\beq \label{wdw}
\bar{\gamma}{}^{IJ} 
 \Big( \der_\mu +   \frac{1}{4} \omega_{\mu KL}\bar{\gamma}^{KL} 
  - 
  \omega_{\mu}{}_{M}{}^{K}\omega_{\beta}{}^{ML} 
  \frac{\der}{\der \omega_{\beta}{}^{KL}} 
  \Big) 
\frac{\der}{\der \omega_{\mu}{}^{IJ}}   
   \Psi 
       = - 
       \lambda \Psi ,     
\eeq
where  $\lambda:= \frac{\Lambda}{(\hbar\kappa\ka)^2}$ is a dimensionless constant which combines three different scales: cosmological, Planck, and 
the 
 UV scale $\varkappa$ introduced by precanonical quantization. 
 
 The fact that all physical constants have   been 
 melted into 
 a single 
 dimensionless constant $\lambda$, which 
 appears as an eigenvalue 
 of the operator in the l.h.s. of eq.~(\ref{wdw}),  
 suggests that the 
 latter, and the theory of quantum gravity derived from precanonical 
 quantization, 
  may correspond to a statement of purely mathematical nature 
 concerning the sections of the Clifford bundle over the bundle 
 of spin connections  over space-time. 
 Surprisingly, eq. (\ref{wdw}) is invariant with respect to the scale transformation $x\!\rightarrow\! a x, \, \omega\!\rightarrow\! a^{-1}\omega $,  
 so that 
 it knows nothing about the Planck scale, 
 where the effects 
of quantum gravity are commonly expected to 
 take place.  
Probably, it is an interaction of gravity with the matter fields, which we  
neglected here,  that (re-)introduces the physical scales.   
 
 
\newcommand{\betab}{\bar{\gamma}{}^0}  
 The scalar product of precanonical wave functions is given by 
\beq \label{ovm}
\left\langle \Phi | \Psi \right\rangle 
:=  \Tr \int  \Phib \, \what{[\rd\omega]}_{} \Psi, \quad 
\what{[d\omega]} 
 =\ri^{\frac12 n(n+1)-1}
 \hat{{\mathfrak e}}{}^{- n(n-1)}\prod_{\mu, I<J} \rd \omega_\mu^{IJ},
\eeq 
where $\Psib\!:=\!\betab\Psi^\dagger\betab$ 
is the complex conjugate and reverse of $\Psi$,\!\cite{snygg} 
 and  $\what{[\rd\omega]}$ 
 with {$\hat{\mathfrak{e}}{}^{-1}\!=\! 
 \frac{1}{n!} \epsilon^{I_1...I_n}\epsilon_{\mu_1...\mu_n} 
\hat{e
}{}^{\mu_1}_{I_1} ... \hat{e}{}^{\mu_n}_{I_n}$}  
 is a Misner-like\cite{misner} 
 diffeomorphism invariant generalized-Hermitian 
 (in the sense $\what{[\rd\omega]}\!=\!\overline{\what{[\rd\omega]}}$)  
 operator-valued measure on the 
fibers of the bundle of spin connections over  space-time.
 In general, $\Tr \big (\Psib\Psi\big)$ 
 on a pseudo-Euclidean Clifford algebra 
 is not positive-definite. 
 However, on the subspace of functions 
 $\Phi^{\pm}\!:=\!(1\pm \betab) \Psi (1\pm \betab)$, 
 it coincides with the positive definite Frobenius inner product: 
 $\Tr \big(\overline{\Phi^{\pm}} \Phi^{\pm}\big)\!=\! 
 \Tr \big({\Phi^{\pm}}{}^\dagger {\Phi^{\pm}}\big)\! 
 =:\!\|\Phi^{\pm}\|^2$. 
 
\newcommand{\oldtextaba}{However, on the subspaces of functions 
 $\Phi^{++}$ and $\Phi^{--}$: 
 $\Phi^{\pm\pm}:= (1\pm \betab) \Psi (1\pm \betab)$,   
 the corresponding inner product on 
 the %
 Clifford algebra:   
 $\Tr \big({\Phi^{\pm\pm}}{}^\dagger {\Phi^{\pm\pm}}\big)$
 $=:||\Phi^{\pm\pm}||^2$, 
 coincides with the positive definite Frobenius inner product 
 $\Tr \big({\Phi^{\pm\pm}}{}^\dagger {\Phi^{\pm\pm}}\big)$
 $=:||\Phi^{\pm\pm}||^2$, 
 so that one can attempt to define the physical 
 Hilbert space using this fact.} 



 The normalizability of precanonical wave functions 
 according to the above scalar product 
 (and quite independently from its details) 
 requires $\hat{{\mathfrak e}}{}^{- \half n(n-1)}\Psi$ 
 to  vanish~at large $\omega$-s. 
 This essentially implies 
 the quantum avoidance of curvature \mbox{singularities} (large $\omega$-s)   
 in the sense that the probability  density 
of observing the regions of space-time with 
an extremely high curvature  vanishes 
as a consequence of the normalizability of 
 the 
 precanonical~wave
 function of quantum gravity. 
 Note, however, that 
 in spite of its plausibility, this resolution 
 to the singularity problem 
 depends on the actual existence 
of the properly normalized solutions of eq.~(\ref{wdw}), which is 
 not~yet proven. 
Moreover, 
 the argument 
 based on the normalizability 
  of precanonical wave functions 
  in its present form 
 ignores 
 interactions with matter, 
 the intricalities 
 related to the indefiniteness of 
{$\Tr (\Psib\Psi)$},   
and the 
gauge fixing, i.e. the choice of~the coordinate 
 systems/local 
orientations of vielbeins on average,   
which is 
important for the extraction of the 
physically relevant information from the solutions~of~(\ref{wdw}). 

%
%

Eq.~(\ref{wdw}) can be seen as a variant of a generalized 
confluent hypergeometric equation of $n$ 
matrix (Clifford-valued) variables $Z_\mu:= \omega^{IJ}_\mu \bar{\gamma}_{IJ}$.  
To my knowledge, the theory 
of such equations is not yet 
 sufficiently developed in mathematics,  
so that even the proof of the existence of properly normalized  solutions, 
which is essential for the present formulation 
of quantum theory of gravity 
 to be physically viable and mathematically well-defined,   
 is a challenging problem. 
 %
 However, 
   in the regime when the term $\omega\omega\der_\omega\der_\omega$ in 
 (\ref{wdw})
can be neglected,    
we obtain a more tractable equation 
\beq
\left( \ri k_\mu \der_{Z_\mu} + 
\mbox{$\frac14$} Z_\mu \der_{Z_\mu} +\lambda \right)\Phi =0 , 
\eeq
where $\Psi(x,\omega) = \re^{\ri k_\mu x^\mu} \Phi(Z_\mu)$ 
and 
 the ordering ambiguity in 
 the 
 $Z\der_Z$ term 
 is accounted for in a re-definition of $\lambda$. 
By separating the variables $Z_\mu$, so that $\Phi(Z)= \Pi_\mu \Phi(Z_\mu)$ 
and $\lambda = \Sigma_\mu c_\mu$, 
we obtain the following equation for each $\mu$:
\beq
(\ri k\der_Z + 
\mbox{$\frac14$} Z\der_Z)\Phi + c\Phi = 0 .
\eeq
Its solution 
(written in terms of the formal  fractional power of the matrix 
in the square brackets\cite{fract})  
 is\footnote{
    A solution similar to (\ref{sol1}) was obtained also 
 in the cosmological context\cite{my-vielbein}{}$^{c,}${}\cite{mg14-cosmol}, 
 where the  $\omega \omega \der_\omega\der_\omega$ term 
 in (\ref{wdw}) vanishes identically because 
 of the cosmological symmetries of the problem, 
 which leave non-vanishing only one independent component of 
 the spin connection.}
\beq \label{sol1}
\mbox{\normalsize $\Phi \sim [4\ri k + Z]^{-4c}. $}
\eeq 
  %
 %
 %
 
 The corresponding probability density 
 of spin connections given by $\|\Phi\|^2$
 behaves as  $\, \sim\!\! \|Z\|^{-8c}$ at 
  $|k|\!\ll\!\|Z\|$, 
  and its 
 normalizability with the measure in (\ref{ovm}) 
 restricts the admissible values of $c$.  
  The above heavy-tailed power law asymptotic behavior 
 is characteristic of 
  L\'evy processes, 
  with the exponent 
  known to be related to the fractal dimensionality 
  of the trajectories  of 
   L\'evy flights\cite{levy} in the target spaces,  
   which are fibers of the bundle of spin connections here.  
A L\'evy flight on each fiber of the bundle means 
a random spin connection field over  space-time 
  with a generally non-Gaussian distribution defined by 
  the 
  precanonical wave function. The properties of this 
  omnipresent 
  random  field will be reflected in the properties 
 of test particles propagating in space-time, 
 their trajectories, spatio-temporal distribution and correlations, 
 which can be potentially observable. The fractality of the 
 L\'evy flights in the 
  target space of spin connections  
  may be translated to the effective 
 fractality of 
  space-time itself as one of the consequences of quantum 
 gravity. 
  We hope to explore the details of that in the future. 
  

\newcommand{\texttosave}{
********
 
with the index of stability  $\mu = 8c -1D??$ 
when $0\leq \mu \leq 2$ (see Ref.~\refcite{west82}).
This corresponds to the fractal dimensionality $\mu$ of the trajectories 
of the corresponding stochastic process,  which 
in our context translates to the fractal dimensionality of the 
space-time with a 
 random distribution of spin connection 
 determined by 
 the 
  precanonical wave function.  

*********

 Let us note that a solution similar to (\ref{sol1}) was obtained also 
in the cosmological context\cite{my-vielbein}{}$^{c,}${}\cite{mg14-cosmol}, 
where the  $\omega \omega \der_\omega\der_\omega$ term 
in (\ref{wdw}) vanishes identically because 
of the cosmological symmetries of the problem 
leaving only one independent component of the spin connection non-vanishing.
 } 


 In the opposite regime, 
 when   
the term $\omega \omega \der_\omega \der_\omega$ in (\ref{wdw})
dominates (and the wave numbers $k_\mu$ are not too high), 
eq. (\ref{wdw}) reduces to a matrix generalization of 
a 
simple PDE:\, $u^2\Phi''(u) - \lambda\Phi(u)=0$, whose power law 
solutions $\Phi\! \sim\! u^{\frac12(1\pm\sqrt{4\lambda+1})}$ 
 again 
 point to a restriction on the admissible values of 
 the cosmological constant in 
 $\lambda$ from 
the normalizability of the wave function at large $\omega$-s,   
 and to  
a heavy-tailed L\'evy-type behavior at large $\omega$-s, 
albeit with a different fractal dimensionality 
of the 
 L\'evy flights in this regime. 



The Green functions of (\ref{wdw}) are 
 transition amplitudes 
between 
the 
values of spin connections at different points:  
  {$\langle \omega,x |\omega',x' \rangle$}.  
They describe a quantum space-time geometry which generalizes 
 the 
 classical  geometry formulated in terms of smooth spin connection 
fields $\omega(x)$. By noticing 
 an 
 analogy with the statistical 
hydrodynamics approach to turbulence,\cite{turbulence}  
 which replaces the description in terms of 
 the smooth velocity fields $\mathbf{v}(\bx)$ 
 with  the statistical description in terms of 
 the multi-point velocity correlators,  
           we can speak of the picture of quantum gravity 
 derived from precanonical quantization 
 as a {\em space-time turbulence} 
 or, 
 in the mathematical context of the present formulation, 
 a {\em spin connection foam}. In fact, this picture is even 
 closer to the original Wheeler's intuition about the 
 \mbox{``space-time foam"}\nopagebreak  
 { }than his quantum geometrodynamics based on the Wheeler--De Witt equation 
 and the notion of an infinite-dimensional superspace of all metrics. 
 
 Note also that the metric tensor in the present formulation is 
  an operator: 
$$\what{g}{}^{\mu\nu}=-\hbar^2 \ka^2\kappa^2 
\eta^{IJ}\eta^{KL}\frac{\der^2}{\der\omega^{IK}_\mu \der\omega^{JL}_\nu}.
$$
 Hence, the 
 intervals $\what{ds}{}^2 \!=\! \what{g}{}_{\mu\nu} dx^\mu dx^\nu$ 
 are operator-valued.  This  makes the notion of the distances between points,  
   and hence 
   the operational notion of a point of 
 the 
 physical space-time, 
 fuzzy 
  due to the quantum nature of 
   the 
   space-time geometry. 
 In this way the description of quantum geometry of space-time
 derived from the precanonical quantization of general relativity 
 complements the 
 ideas about 
 quantum space-time   
 being discussed within the framework of LQG, string theory  and non-commutative geometry.

\section{Possible connections with LQG}

A first discussion of the constraints in 
  the Ashtekar formulation 
 using a version of the 
 DW (multisymplectic) Hamiltonian theory 
 applied to the vielbein Einstein-Palatini Lagrangian (\ref{lagr}) 
 can be found in the paper  by Esposito e.a.\cite{esposito}
 
In spite of the obvious differences of our precanonical 
 approach 
from the LQG programme, 
which (i) uses a (3+1)-splitting 
 vs. our explicitly space-time symmetric approach, 
(ii) heavily relies on the specifics of four-dimensional 
 globally hyperbolic space-times  
vs. our intention 
to  develop a formalism potentially applicable
in any number of dimensions and any signature of space-time,
(iii) uses functionals and functional derivative operators 
vs. our use of functions and partial derivative operators,  
etc., there are some striking similarities as well, one of them 
being the emergence, after Hamiltonization,  
of a formulation based on the connections,  with the 
vielbeins, or the densitized inverse triads/dreibeins in the Ashtekar formulation,  
represented as differential operators with respect to the connections. 

Whereas our approach to quantization of gravity 
is based on the fundamental brackets  
in the Weyl subalgebra in the subspace of Hamiltonian $0$- and $(n$--$1)$-forms, 
our construction of the Poisson-Gerstenhaber brackets in field theory\cite{my-pbr} 
also offers another, yet unexplored 
opportunity based on the Dirac bracket between the spin connection 1-form 
$\omega_\alpha^{IJ}dx^\alpha$ and 
its conjugate $(n$--$2)$-form of polymomenta: 
${\mathfrak p}^\mu_{\omega_\nu^{KL}}  \varpi_{\mu\nu}$, where
$ \varpi_{\mu\nu}:= 
\imath_{{}_{\der_\mu}}\imath_{{}_{\der_\nu}} \varpi$:  
 \beq \label{asht} 
\pbr{\omega_\alpha^{IJ}dx^\alpha}{{\mathfrak p}^\mu_{\omega_\nu^{KL}}  \varpi_{\mu\nu}}{}^{\rm D} 
 = \pbr{\omega_\alpha^{IJ}dx^\alpha}{{\mathfrak p}^\mu_{\omega_\nu^{KL}}  \varpi_{\mu\nu}}
\sim n\, \delta^{[I}_K \delta^{J]}_L  . 
\eeq    
This bracket 
  is a gravitational analogue of 
  the bracket between the potential 1-form 
$\mathsf{A}=A_\mu dx^\mu$ and the $(n$--$2)$-form of the 
dual field strength in Electrodynamics,
$*\mathsf{F}=*\half F_{\mu\nu}dx^\mu\we dx^\nu$,  
  found in Ref.~\refcite{my-pbr}  
 (see also Refs.~\refcite{AF1,AF2}).
After a 
 (3+1)-splitting,  
 a restriction to the initial data surfaces and 
 integration over them  
(cf. Refs.~\refcite{my-pbr,AF1}),  
the bracket can be related to the fundamental Poisson 
bracket underlying the Ashtekar formulation in (3+1)-dimensions, 
and the forms involved in (\ref{asht})
can be used to construct the holonomy-flux variables underlying the LQG 
approach (see e.g. Refs.~\refcite{ashtekar} for a review). 
However, 
unlike the subalgebra of $0$- and $(n$--$1)$-forms in 
(\ref{dbr11})--(\ref{dbr13c}), 
  the set of $(n$--$2)$- and $1$-forms is not closed 
  with respect to the Poisson-Gerstenhaber bracket operation, 
  hence the closure of the subalgebra 
   involves 
  forms of all the degrees 
  $\leqslant\! (n$--$2)$.
   

A further understanding of possible connections of 
precanonical quantization with LQG in (3+1) dimensions 
requires an inclusion of the Holst term\cite{holst} 
in the Einstein-Palatini Lagrangian (\ref{lagr})  
 in order to incorporate the Barbero-Immirzi \mbox{parameter,}  
 that does not, however, appear to be  necessary within our approach. 
 %
Interestingly, this property is shared with a 
multidimensional generalization of the LQG-type 
connection formulation of pure gravity 
 discussed by Bodendorfer e.a.\cite{thiemann} 
Hence, it is likely that more connections can be found 
between that formulation and the precanonical approach 
of the present paper. 



\end{document}